\documentclass[
    ,final            
  ]
  {aipproc}

\layoutstyle{6x9}

\begin{document}

\title{Elliptic Flow and Shear Viscosity within a Transport Approach from RHIC to LHC Energy}


\author{S. Plumari}{
  address={Department of Physics and Astronomy, University of Catania, Via S. Sofia 64, I-95125 Catania},
  altaddress={Laboratorio Nazionale del Sud, INFN-LNS, Via S. Sofia 63, I-95125 Catania}
}

\author{V. Greco}{
  address={Department of Physics and Astronomy, University of Catania, Via S. Sofia 64, I-95125 Catania},
  altaddress={Laboratorio Nazionale del Sud, INFN-LNS, Via S. Sofia 63, I-95125 Catania}
}

\begin{abstract}
We have investigated the build up of anisotropic flows within a parton cascade approach 
at fixed shear viscosity to entropy density $\eta/s$ to study the generation of collective flows 
in ultra-relativistic heavy ion collisions. 
We present a study of the impact of a temperature dependent $\eta/s(T)$ on the generation 
of the elliptic flow at both RHIC and LHC.
Finally we show that the transport approach, thanks to its wide validity range, is able to describe 
naturally the rise - fall and saturation of the $v_2(p_T)$ observed at LHC. 
\end{abstract}

\maketitle


\section{Introduction}
The RHIC program at BNL has shown that the azimuthal asymmetry in momentum space,
namely the elliptic flow $v_2$ , is the largest ever seen in HIC suggesting that 
an almost perfect fluid with a very small shear viscosity to entropy density ratio, 
$\eta/s$, has been created \cite{STAR_PHENIX}. The first measurement at LHC in Pb+Pb 
at $2.76 \, TeV$ \cite{ALICE_2011} shows that the integrated elliptic flow as a function of 
collision energy increase of about $30 \%$ compared to the flow measured at RHIC energy 
of $200 \, GeV$, while the $v_2(p_T)$ measured at LHC comparted to that of RHIC does not change 
indicating an increase in the average transverse momentum. It remains to be understood if this 
means an equal $\eta/s$ of the formed plasma or it is the result of different initial conditions 
and possible larger non-equilibrium effects.

The most common approach to study viscous correction is viscous hydrodynamics at second order in 
gradient expansion according to the Israel-Stewart theory \cite{Romatschke:2007mq, Molnar_cascade, Heinz}. 
This approach has been implemented to simulate the RHIC collisions providing 
an upper bound for $\eta/s \leq 0.4$. Such an approach, apart 
from the limitation to 2+1D simulations, has the more fundamental problem of a limited range of 
validity in $\eta/s$ and in the transverse momentum $p_T$. In these proceedings we discuss results 
within the relativistic transport approach that has the advantage to be a 3+1D approach
not based on a gradient expansion in viscosity that is valid also for large 
$\eta/s$ and for out of equilibrium momentum distribution allowing a reliable 
description also of the intermediate $p_T$ range. In this $p_T$ region viscous 
hydrodynamics breaks its validity because the relative deviation of the equilibrium 
distribution function $\delta f / f_{eq} $ increases with $p_T^2$ becoming large 
already at $p_T \geq 3T \sim 1 GeV$.
In the following we will show the results obtained with a parton cascade approach where 
the EoS is fixed to the one of a free massless gass $\epsilon-3P=0$ and the mean free 
path $\lambda$ is finite. A more quantitative comparison with the experimental data 
would require the inclusion of mean field dynamics associate to an equation of state 
$P(\epsilon)$ according to lQCD results \cite{Wuppertal_Budapest}.
A first step in this direction has been discussed in \cite{Plumari:NJL,Plumari:HotQuarks} 
while the implementation of a quasi-particle model \cite{Plumari:qp_model} in the 
contest of transport theory is in progress along lines similar to \cite{Cassing}.

\section{The Parton Cascade at fixed $\eta/s$}
Our approach is a $3 + 1$ dimensional Montecarlo cascade \cite{greco_cascade} for on-shell 
partons based on the stochastic interpretation of the transition rate discussed in 
Ref. \cite{Greiner_cascade}.
In kinetic theory under ultra-relativistic conditions the shear viscosity can be 
expressed as $\eta=(4/15) \rho <p> \lambda$
with $\rho$ the parton density, $\lambda=[\rho \sigma_{tr}]^{-1}$ the mean free path 
and $<p>$ the average momentum. Therefore considering that the entropy density for a 
massless gas is $s=\rho (4 - \mu/T)$, $\mu$ being the fugacity, we get:
\begin{equation}
\eta/s=\frac{4}{15}\frac{<p>}{\sigma_{tr}\rho (4 - \mu/T)} \label{eq:eta_s}
\end{equation}
where $\sigma_{tr}$ is the transport cross section. In our approach we solve the relativistic 
Boltzmann equation with the constraint that $\eta/s$ is fixed during the dynamics of the
collisions in a way similar to \cite{Molnar_1}
but with an exact local implementation a more detailed discussion of the method is in 
\cite{greco_cascade}. In fact fixing $\eta/s$ we can evaluate locally in space and time the strength of 
the cross section $\sigma_{tr}(\rho,T)$ needed to have $\eta/s$ at the wanted value 
by mean of the following formula:
\begin{equation}
\sigma_{tr}=\frac{4}{15}\frac{<p>}{\rho (4 - \mu/T)}\frac{1}{\eta/s} \label{eq:eta_s2}
\end{equation}
This approach is equivalent to have a total cross section of the form 
$\sigma_{Tot}=K(\rho, T) \sigma_{pQCD} > \sigma_{pQCD}$ where $K$ takes into account the non perturbative 
effects responsible for that value of viscosity. Note that this approach have been shown to recover 
the viscous hydrodynamics evolution of the bulk system \cite{Molnar_cascade, greco_cascade}, but 
implicitly assume that also high $p_T$ particles collide with largely nonperturbative cross section. 
We show here that both at RHIC and LHC there are signatures of the disappearence of the large non 
perturbative physics with increasing $p_T$.

\subsection{Effect of temperature dependent $\eta/s$(T)}

\begin{figure}
  \includegraphics[height=.21\textheight]{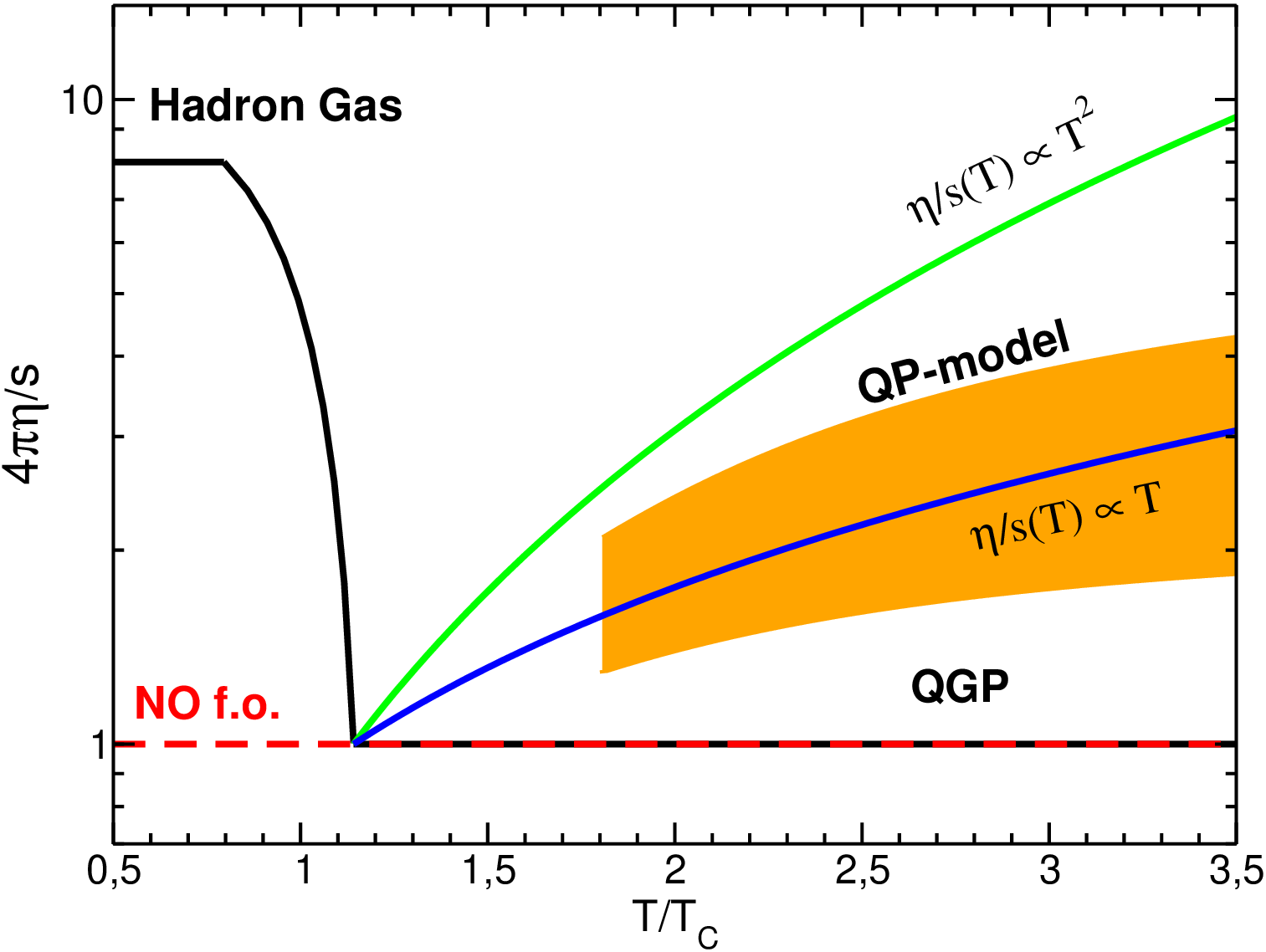}
  \caption{Different temperature dependent parametrizations for $\eta/s$. 
  The orange area take into account the quasi-particle model predictions for $\eta/s$ \cite{Plumari:qp_model}.}
  \label{Fig:etas_T}
\end{figure}
In our calculation the initial condition are longitudinal boost invariant
with the initial parton density $dN/d\eta(b=0)=1250$ at RHIC and $dN/d\eta(b=0)=2250$ at LHC.
The partons are initially distributed in coordinate space according to the 
Glauber model while in the momentum space at RHIC (LHC) the partons with $p_T \leq p_0=2 GeV$ 
($p_T \leq p_0=4 GeV$) are distributed according to a thermalized spectrum with a maximum 
temperature in the center of the fireball of $2 T_C$ ($3.5 T_C$), 
while for $p_T > p_0$ we take the spectrum of non-quenched minijets according to standard 
NLO-pQCD calculations.
We also start our simulation at the time $t_0 = 0.6 fm/c$ at RHIC and $t_0=0.3 fm/c$ at LHC.

\begin{figure}
  \includegraphics[height=.24\textheight,width=.7\textwidth]{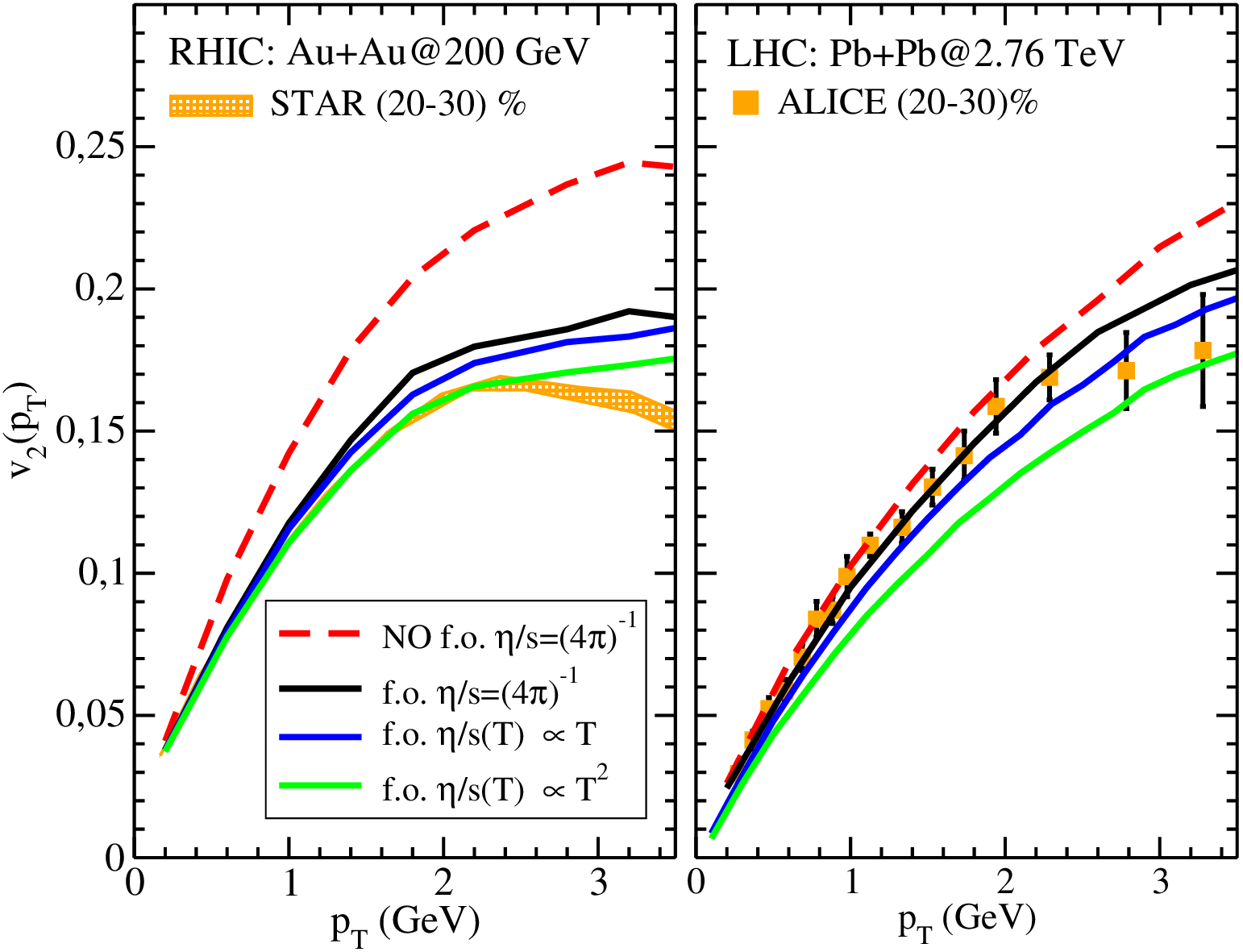}
  \caption{Differential elliptic flow $v_2(p_T)$ at mid rapidity for $20\%-30\%$ collision centrality. On the left panel, the orange band indicate RHIC results measured by STAR and the orange points on the right panel are the LHC results measured by the ALICE collaboration, data taken by \cite{ALICE_2011}. The red dashed line is the calculation with $4\pi \eta/s = 1$ during all the evolution of the fireball and without the freeze out condition, while the black blue and green lines are calculations with the inclusion of the kinetic freeze out and with $4\pi\eta/s=1$, $4\pi\eta/s \propto T$ and $4\pi\eta/s \propto T^2$ in the QGP phase respectively.}
  \label{Fig:v2_etasT}
\end{figure}

In order to study the effect of the kinetic freezeout on the generation of the elliptic flow 
we have performed two calculations one with a constant $4\pi\eta/s=1$ during all the evolution 
of the system (red dashed line of Fig.\ref{Fig:etas_T}) the other (shown by black solid line 
in Fig.\ref{Fig:etas_T}) with $4\pi\eta/s=1$ in the QGP phase and an increasing 
$\eta/s$ in the cross over region towards the estimated value for hadronic matter $4 \pi \eta/s = 8$ 
\cite{etaS_hadronic}. Such an increase allows for a smooth realistic realization of the kinetic 
freeze-out.
In Fig. \ref{Fig:v2_etasT} it is shown the elliptic flow $v_2(p_T)$ at mid rapidity for 
$20\%-30\%$ centrality for both RHIC Au+Au at $\sqrt{s}=200 GeV$ (left panel) and LHC 
Pb+Pb at $\sqrt{s}=2.76 TeV$ (rhigh panel).
As we can see at RHIC energies, left panel of Fig. \ref{Fig:v2_etasT}, the $v_2$ is sensitive to 
the hadronic phase and the effect of the freeze out is to reduce the $v_{2}$ of about of $25 \%$, 
from red dashed line to black solid line in left panel of Fig. \ref{Fig:v2_etasT}. 
For the $p_{T}$ range shown we get a good agreement with the experimental data for a minimal 
viscosity $\eta/s \approx 1/(4\pi)$ once the f.o. condition is included.
At LHC energies, right panel of Fig. \ref{Fig:v2_etasT}, the scenario is different, we have 
that the $v_2$ is less sensitive to the increase of $\eta/s$ at low temperature in the hadronic 
phase. The effect of large $\eta/s$ in the hadronic phase is to reduce the $v_2$ by less than 
$5 \%$ in the low $p_T$ region, from red dashed line to the black solid line in right panel 
of Fig. \ref{Fig:v2_etasT}. 
This different behaviour of $v_2$ between RHIC and LHC energies can be explained looking 
at the life time of the fireball. In fact at RHIC energies the life time of 
the fireball is smaller than that at LHC energies, $5 fm/c$ at RHIC against the about $10 fm/c$ 
at LHC. Therefore at RHIC the elliptic flow has not enough time to fully develop in the QGP phase. 
While at LHC we have that the $v_2$ can develop almost completely because the fireball 
spend more time in the QGP phase.

Due to this large life time of the fireball at LHC and the larger initial temperature 
is interesting to study the effect of a temperature dependence in $\eta/s$.
In the QGP phase $\eta/s$ is expected to have a minimum of $\eta/s \approx (4\pi)^{-1}$ close to 
$T_{C}$ as suggested by lQCD calculation \cite{lQCD_eta}. While at high temperature quasi-particle 
models seems to suggest a temperature dependence of the form $\eta/s \sim T^{\alpha}$ with 
$\alpha \approx 1 - 1.5$ \cite{Plumari:qp_model}. To analyze these possible scenarios
for $\eta/s$ in the QGP phase we have considered two different situation one with a linear dependence 
$4\pi \eta/s=T/T_0=(\epsilon/\epsilon_0)^{1/4}$ (blue line) and the other one with a quadratic 
dependence $4\pi \eta/s=(T/T_0)^2=(\epsilon/\epsilon_0)^{1/2}$ (green line) where 
$\epsilon_0=1.7 GeV/fm^3$ is the energy density at the beginning of the cross over regions 
where the $\eta/s$ has its minimum, see Fig.\ref{Fig:etas_T}.
At RHIC energies the $v_2$ is essentially not sensitive to the dependence of $\eta/s$ on temperature 
in the QGP phase, see the blu and green lines in the left panel of Fig. \ref{Fig:v2_etasT}.
However the effect on average is to decrease the value of $v_2$ but at low $p_T < 1.5 GeV$ 
the $v_2(p_T)$ appears to be insensitive to $\eta/s(T)$ while a quite mild dependence appears at higher 
$p_T$ where however the transport approach tends always to overpredicted the elliptic flow 
observed experimentally.
At LHC energies the build-up of $v_2$ is more affected by the $\eta/s$ in the QGP phase and 
on average it is reduced of about a $20 \%$. In any case still a strong temperature 
dependence in $\eta/s$ has a small effect on the generation of $v_2$ 
we found that with a constant or at most linearly dependent on T $\eta/s$ the transport 
approach can describe the data at both RHIC and LHC at least up to $p_T \sim 2 GeV$. However 
the transport approach should keeps its validity also at higher $p_T$, but as previously said, the 
agreement with data seem to weaken at $p_T > 2 GeV$ both at RHIC and LHC. We discuss the underlying 
reasons in the next section.


\subsection{Impact of high $p_{T}$ partons on $v_{2}$}
In our approach we have that $\sigma_{Tot}=K \sigma_{pQCD}$ therefore we have 
large cross section independently of the $p_T$ of the colliding particles. 
But we know that particles with high energies should collide with the pQCD cross section. 
In order to take into account the proper scattering cross section for hard collisions we 
extend our previous approach allowing for a $K$ factor that depends on the invariant 
energy of the collision $K(s)$ which gives the connection between the non pertubative 
interacting bulk and the asymptotic pQCD limit. We choose this function in such a way that at 
high energies $K(s) \to 1$ and we get the correct assumed pQCD limit. 
For the function $K(s)$ we choose an exponential form  $K(s/\Lambda^2)=1 + \gamma \, e^{-s/\Lambda^2}$, 
where $\Lambda$ is a scale parameter that fix the energy scale at which the pQCD 
behaviour begins to be reached. While $\gamma$ plays the same role of $K$ in the 
old calculations and it is determined again in order to keep fixed locally the $\eta/s$. 
Therefore we can repeat the same procedure as described in the previous section but now 
with $\sigma_{Tot}=K(s/\Lambda^2) \sigma_{pQCD}$. 
\begin{figure}
  \includegraphics[height=.25\textheight]{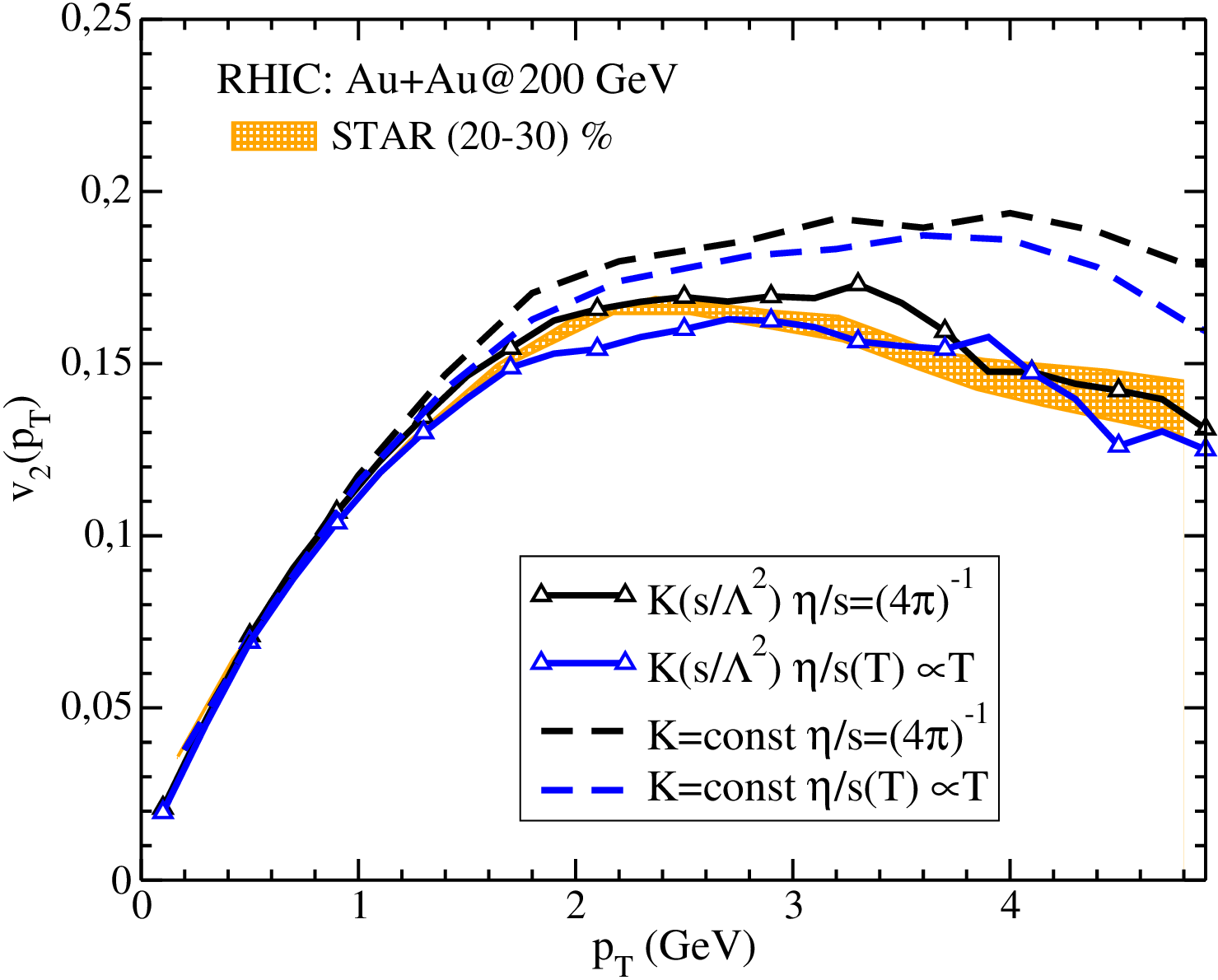}
  \hskip1cm
  \includegraphics[height=.25\textheight]{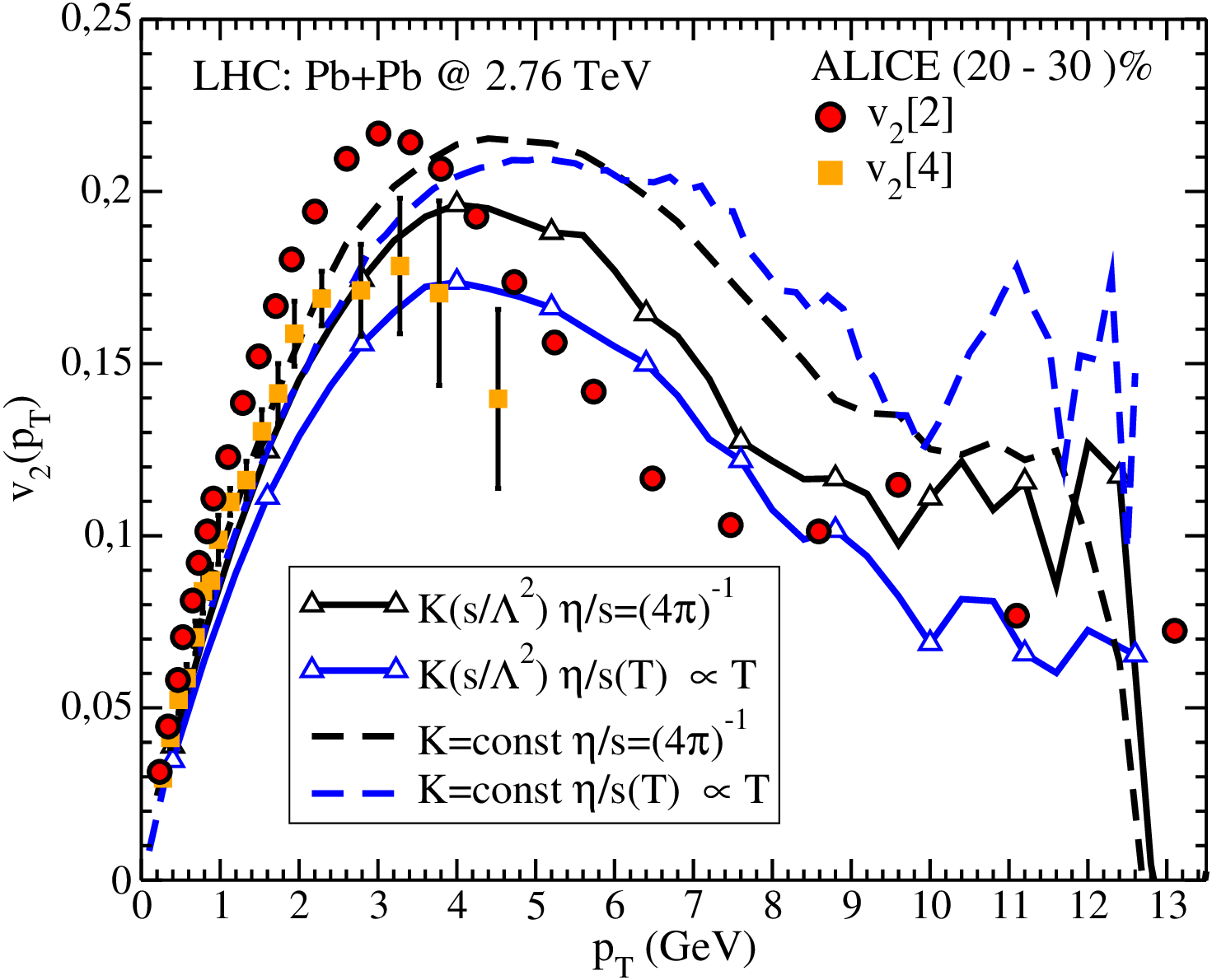}
  \caption{Left: $v_2(p_T)$ at mid rapidity for $20\%-30\%$ collision centrality at RHIC. The dashed lines are the calculations with $K=const$ and for $4\pi\eta/s=1$ and $4\pi\eta/s \propto T$ with f.o. respectively for black and blue curves while the solid lines are the same but with $K(s/\Lambda^2)$. Right: $v_2(p_T)$ at mid rapidity and for $20\%-30\%$ collision centrality at LHC with the same legend, data taken from \cite{ALICE_2011}.}
  \label{Fig:v2_Ks_etas}
\end{figure}
Due to its physical meaning we of course expect $\Lambda$ to be greater than $2 GeV$, in particular 
we have performed different calculation for different value of $\Lambda$ and we have 
obtained that for $\Lambda > 4 GeV$ the elliptic flow becomes less sensitive to the 
value of the parameter $\Lambda$. Specifically in our calculation we have considered 
the value $\Lambda = 4 GeV$.
As we can see at RHIC energies, left panel of Fig. \ref{Fig:v2_Ks_etas}, we have that 
$K(s/\Lambda^2)$ does not affect at all the $v_2(p_T)$ for $p_T < 2 GeV$, in other words 
high $p_T$ parton at RHIC energies does not affect the generation of the $v_2$ of 
the bulk. 
Instead we have a reduction of the $v_2$ for $p_T > 3 GeV$ and with the inclusion 
of $K(s/\Lambda^2)$ the $v_2$ becomes a decreasing function of $p_T$ for $p_T > 3 GeV$
in perfect agreement with what is  observed experimentally (orange band).
In Fig. \ref{Fig:v2_Ks_etas} (right) we compare in a large range the $v_2(p_T)$ at LHC 
energy with (solid) and without (dashed) the inclusion of an energy dependent $K$ 
factor and for two T dependence of the $\eta/s$. We notice that the two sets of experimental 
data refer to different method of $v_2$ measurements, namely $v_2[2]$ (circle) and 
$v_2[4]$ (square) and our theorethical results should be compared to $v_2[4]$ because 
event-by-event fluctuations are not considered.
As we can see at LHC energies the $v_2$ is sensitive to $K(s/\Lambda^2)$ already at 
$p_T \approx 1.5 GeV$ quite lower than the RHIC case, in other words the many high 
$p_T$ partons that we have at LHC energies affect the generation of the $v_2$ of the bulk. 
Similar results we have when we include a $\eta/s(T)$ in the QGP phase.

At low $p_T$ the raise of the $v_2$ is an effect of a strong interacting fluid with a very 
small viscosity. In this regime we have that particles with low $p_T$ interact non perturbatively
with large cross sections and therefore we get a description in agreement with hydrodynamics.
With the increase of the $p_T$ of the partons the pQCD limit begins to be important 
and for $p_T > 3 - 4 GeV$ the elliptic flows starts to be a decreasing function of $p_T$.
The disappearance of the non perturbative effect significantly affects the $v_2(p_T)$ making faster 
and stronger ($\sim 20 - 25 \%$) the fall in the elliptic flow in the range $3 GeV < p_T < 8 GeV$. 
Finally for $p_T > 8 GeV$ in our calculation seems to appear the 
saturation of the $v_2$ similarly to the experimental data and typical of a path-lengh mechanism
as in jet quenching models \cite{Scardina}. An analysis with better 
statistics is required. In this range of $p_T$ the only effect is that given by 
the pQCD limit.


\section{Conclusion}
We have investigated within a transport approach at fixed $\eta/s$ the effect of a 
temperature dependent $\eta/s$ at RHIC and LHC energies.
At RHIC we have seen that the elliptic flow is more sensitive to the kinetic freeze out 
(hadronic phase) and still of $25 \%$ of $v_2$ depends on it.
At LHC we get an opposite effect, nearly all the $v_2$ comes from the QGP phase and the 
$\eta/s$ of the hadronic phase is irrilevant. 
We get for both at RHIC and LHC a good agreement with the data when the ratio $\eta/s \approx 1/(4\pi)$ 
or a linear T dependence is considered, in general we observe not a large sensitivity of 
$v_2$ to the T dependence in $\eta/s$. 
Furthermore we have seen that at LHC the large ammount of particle with $p_{T} > 4 GeV/c$ 
interacting nearly perturbatively cannot be neglected.  
The interplay between perturbative and non-perturbative behaviour seems to have an important 
effect on the generation of $v_2$ at intermediate $p_T$ and it could explain the rapid raise 
and fall of $v_{2}(p_{T})$ in $0<p_{T}<8 GeV/c$ shown in the experiments.


\begin{theacknowledgments}
This work was supported in part by funds provided by the Italian Ministry of Education, 
Universities and Reserach under the Firb Research Grant RBFR0814TT.
\end{theacknowledgments}



\bibliographystyle{aipproc}   





\end{document}